# NEAR-RESONANT INTERACTIONS OF THE MAGNETOROTATIONAL INSTABILITY IN THIN KEPLERIAN DISCS

*SUB-TITLE:* **NEAR-RESONANT INTERACTIONS OF MRI IN THIN DISCS**

Yuri Shtemler,  Michael Mond, and Edward Liverts

[1]*Department of Mechanical Engineering, Ben-Gurion University of the Negev,*

*P.O. Box 653, Beer-Sheva 84105, Israel*

**ABSTRACT**

The equations that govern the weakly nonlinear near-resonant interaction of a parent magnetorotational instability with two linearly stable daughter modes in thin nonuniform magnetized Keplerian discs are derived. While the equations for the daughter modes resemble their first order classical uniform counterparts, the parent mode is governed by a second order forced Duffing equation. It is demonstrated that the solutions of those three coupled equations exhibit a wide and rich spectrum of dynamical behavior. In particular, it is shown that amplitudes of unstable triads may grow explosively with time, namely reaching infinite values in a finite time. . Paradoxically, explosively unstable near- resonance triads may grow much faster than their strict-resonance counterparts.

# 1. INTRODUCTION

Nonlinear interaction between three plasma waves in infinite media may give rise to explosive instabilities under which the amplitudes of all three interacting waves reach infinite values in a finite time [e.g. Dum and Sudan (1969), Rosenbluth et al. (1969), Rabinovich and Reutov (1973), Craik (1985)]. For strict resonance case this phenomenon has been consistently explained theoretically by Zakharov and Manakov (1973), (1975) [see also Craik (1985), Segur (2008), Kono and Skoric (2010) and references therein], which separate the general resonance interactions of three one-dimensional wave packets in a nondissipative medium into parametrical (decay) and explosive ones. In addition nonzero frequency mismatch stabilizes the explosive instability if the initial amplitudes of interacting waves exceed a threshold proportional to the mismatch module value [Rosenbluth et al. (1969)].

The existence of explosive instabilities in thin axially nonuniform Keplerian discs immersed in poloidal magnetic fields is demonstrated below for small frequency mismatch $\Delta\omega$. The effect of non-zero $\Delta\omega$ on the magneto-rotational decay instability (MRDI) of strict resonance triads is studied in the vicinity of the threshold of the magnetorotational instability (MRI) [Velikhov (1959) and Chandrasekhar (1960), Balbus & Hawley (1991)]. The MRDI arises due to the strict resonant coupling of three eignmodes of thin Keplerian discs: a background MRI (slow Alfv´en-Coriolis (AC)) mode with zero real frequency slightly above its threshold, a stable (fast or slow) AC mode and a stable magnetosonic (MS) mode. It has been investigated recently by the method of two-scale asymptotic expansions in the fast and slow times, which reflect the fast oscillations of both stable daughter AC and MS modes, and the slow growth of the zero-frequency parent MRI mode, respectively [Shtemler et al. (2013), Shtemler et al. (2014)]. The following main differences between thin disc and infinite plasma system have been pointed out: (i) parametric dependence on the radial variable due to smallness of the radial derivatives drop out component of the resonant relation for the radial wave numbers; (ii) the amplitude of the MRI mode is much larger than the consequent amplitudes of the AC and MS modes instead of comparable amplitudes of all the triad's components in the standard case. It has also been also demonstrated that the amplitude equations are not of the standard form of their classical predecessor [Craik (1985)]. Instead of the classical stable Alfv\'en wave in infinite plasma system [Galeev & Oraevskii (1962-1963); Sagdeev and Galeev (1969)], the role of the background wave is currently played by a MRI mode that is slightly above the instability threshold. Since the linear instability threshold is crossed at a zero eigenvalue with multiplicity two, the amplitude of the MRI mode is governed by a second order forced Duffing equation with a cubic nonlinear term and a constant force term. The latter is proportional to the product of the amplitudes of the AC and MS daughter waves at a given initial time. This is in contrast to the standard first-order amplitude equations with quadratic nonlinear terms that are cyclically permutted in amplitudes of the three modes. That Duffing equation decouples from the standard differential equations for the amplitudes of the daughter AC and MS waves due to the constancy of the Duffing force in time for $\Delta\omega = 0$. Summarizing the previous studies for $\Delta\omega = 0$ the three modes have been found to strongly deviate from their respective original linear temporal evolution due to their mutual resonant coupling. Thus, the parent MRI mode is nonlinearly saturated by the MRI-driven MS mode and instead of an exponential growth it exhibits nonlinear bursty oscillations with a constant amplitude. Furthermore, when the resonating daughter wave is a slow AC mode, the linearly stable pair of the daughter AC and MS waves is nonlinearly destabilized and grow exponentially in time by tapping into the MRI energy. If, however, a fast AC mode is one of daughter waves, the amplitudes of all three modes remain bounded as they exchange energy in a manner that resembles the classical decay instability.

The present study is devoted to the dynamics of near-resonance triads. Although most properties of strict-resonance systems are inherited by detuned resonant systems, a small nonzero mismatch is shown to dramatically change their nonlinear behavior in thin discs. In particular, decay instability of triads for $\Delta\omega = 0$ can be replaced by an explosive magneto-rotational instability for near-resonance unstable triads. this can occur due to the deviation of the Duffing's oscillator force from a zero-mismatch constant quantity.

The paper is organized as follows. The physical model is formulated in Section 2. It contains two-time asymptotics for the amplitude equations of the resonant triads, evolution equations for the amplitudes of the near-resonance triads and explosive solutions for unstable triads. Results of simulations for stable and unstable triads are presented in Section 3. Summary and discussion are given in Section 4.

## 2. THE PHYSICAL MODEL FOR THIN KEPLERIAN DISCS.
### 2.1 Two-time asymptotics for amplitude equations of the resonant triads

The AC modes in thin Keplerian discs describe incompressible in-plane perturbations and include two families of eigen-oscillations: the slow-AC waves (the unstable members of that family are the MRI modes) and the stable fast-AC waves. The MS modes describe compressible vertical fluctuations. In the thin disc limit they involve perturbations of the density, pressure and the axial velocity, while the perturbed magnetic field is negligible as the effective wave vector is almost parallel to the external magnertic field. Independent in the linear regime, nonlinear interaction of two AC and one MS modes is the subject of the current study. A detailed discussion of the linear dynamics of thin Keplerian discs may be found in \cite{Shtemler et al. 2014}.

The weakly nonlinear interaction between the above mentioned two daughter waves and a parent MRI mode is analyzed by the method of two-scale asymptotic expansions in the fast and slow times. The latter reflect the fast oscillations of both linearly stable AC and MS modes, and the slow growth of the MRI mode, respectively. For small super-criticality of the background MRI mode, $\gamma$, the following fast and slow times $\bar{\tau}$ and $\tilde{\tau}$, respectively, naturally emerge [Shtemler et al. (2014)]:

$$\bar{\tau} = \omega_a \, \tau, \qquad \tilde{\tau} = \gamma\tau, \qquad \frac{\partial}{\partial \tau} = \omega_a \frac{\partial}{\partial \bar{\tau}} + \gamma \frac{\partial}{\partial \tilde{\tau}}, \qquad (\gamma \ll \omega_a \sim \gamma^0). \tag{1}$$

Here $\tau$ is the dimensionless time; $\omega_a$ are the eigenfrequencies of the AC eigenoscillations [Shtemler et al. (2011)]:

$$\omega_a \equiv \omega_{k,l}(\beta) = \sqrt{\frac{\beta + 6\beta_k + l\sqrt{(\beta+6\beta_k)^2 - 36\beta_k(\beta_k - \beta)}}{2\beta}}, \tag{2}$$

$\ell = -1$ ($\ell = 1$) designate the slow (fast) AC modes, while $k$ denotes the axial quantum number of the discrete AC spectrum ($k \geq 1$); $\beta$ is the ratio of the thermal to magnetic pressure and is a function of the radial coordinate; $\beta_k = k(k+1)/3$ is the threshold beta for the destabilizing $k$-th AC mode. The frequencies of the fast AC mode ($l = +1$) are real for arbitrary $\beta$. The slow AC modes ($l = -1$) give rise to the familiar MRI with $k = 1$. Thus, for $\beta_1 \leq \beta \leq \beta_k$ there are exactly $k$ MRI modes in the system. In the current work, discs with $b$ values slightly above $\beta_1$ are considered such that only the first MRI mode may be excited, and in addition, its growth rate is much smaller than the inverse rotation time. Equation (2) has been obtained by replacing the analytically derived Gaussian isothermal density profile by a hyperbolic distribution [Gammie & Balbus (1994), see also Latter et al. (2010) and Shtemler et al. (2011)]. This provides an effective tool to analytically calculate the eigenfunctions as well as the eigenvalues of the linear stability problem for both the MRI and AC modes [Shtemler et al. (2011)] as well as for the MS modes [Liverts et al. (2012a), Liverts et al. (2012b)]. It has been also shown that the spectrum of the MS modes is stable and continuous under the hyperbolic axial density profile approximation. In that case, taking into account that the real part of the frequency of the MRI is zero, the resonant condition may be fulfilled by any three modes that include the MRI mode, a stable (real frequency) AC mode, and a MS mode. In the more realistic case of Gaussian axial density profile, the spectrum is stable albeit discrete. Non the less, it has been shown that in that case for every stable AC mode, a MS mode may be found whose eigen frequency differs from that of the AC mode by an amount that is of the order of the small parameter that describes the nonlinear interaction (the growth rate in the current work, see examples in Shtemler et al. (2014)).

The solution of the weakly nonlinear problem for the modes that are involved in the resonant triad is represented therefore as the sum of zero, first and higher harmonics in the fast time scale $\bar{\tau}$ [Shtemler et al. (2014)]:

$f(\eta, \tau) = f_0\,(\eta, \tilde{\tau}) + [f_1\,(\eta, \tilde{\tau})e^{-i\bar{\tau}} + \cdots + c.c.],$

$$g(\eta, \tau) = g_0(\eta, \tilde{\tau}) + [g_1(\eta, \tilde{\tau})e^{-i\tilde{\tau}} + \cdots + c.c.], \tag{3}$$

where c.c. denotes complex conjugated terms; $\eta = z/H(\varepsilon r)$ is the stretched axial coordinate $z$, $H(\varepsilon r)$ is the effective height of a Keplerian disc that parametrically depends on radial coordinate $r$, $\varepsilon$ is the characteristic disc aspect ratio; $f$ stands for any of the variables that characterize the AC modes, namely, $b_r$, $ib_\theta$, $iv_r$, $v_\theta$, while $g$ describes the MS modes, namely, $iv_z$, or $\nu$; the subscripts denote the harmonic-number in the fast time; $f_0(\eta, \tilde{\tau}), g_0(\eta, \tilde{\tau})$ and $f_1(\eta, \tilde{\tau}), g_1(\eta, \tilde{\tau})$ are the real amplitudes of the zero-harmonics and the complex amplitudes of the first-harmonics, respectively; the dots represent higher-harmonics in the fast time which are also of higher order in the weakly nonlinear hierarchy. Based on insight gained from the nonresonant case, the following form is assumed for the amplitudes of the zero and first harmonics to lowest order in $g$:

$$f_0(\eta, \tilde{\tau}) = A_0(\tilde{\tau})\hat{f}_0(\eta) + A_0(\tilde{\tau})H_0(\tilde{\tau})\hat{f}_{0,0}(\eta) + [A_1^*(\tilde{\tau})H_1(\tilde{\tau})\hat{f}_{-1,1}(\eta) + c.c.] + \cdots,$$
$$f_1(\eta, \tilde{\tau}) = A_1(\tilde{\tau})\hat{f}_1(\eta) + A_0(\tilde{\tau})H_1(\tilde{\tau})\hat{f}_{0,1}(\eta) + c.c. \ldots,$$
$$g_0(\eta, \tilde{\tau}) = H_0(\tilde{\tau})\hat{g}_{0,0}(\eta) + \ldots,$$
$$g_1(\eta, \tilde{\tau}) = H_1(\tilde{\tau})\hat{g}_1(\eta) + A_0(\tilde{\tau})A_1(\tilde{\tau})\hat{g}_{0,1}(\eta) + c.c. \ldots \tag{4}$$

Recalling that the MS spectrum does not include modes with zero eigenvalues, it may be easily shown that $A_0(\tilde{\tau})$ and $H_0(\tilde{\tau})$ are related by the following algebraic relationship: $H_0(\tilde{\tau}) = A_0^2(\tilde{\tau})$.

To summarize, the scenario that emerges is that of a triad of small eigen perturbations that satisfy the resonant condition, initially coexist without mutual interaction. At that stage, in the absence of nonlinear interaction $A_1(\tilde{\tau})$ and $H_1(\tilde{\tau})$ (the amplitudes of the duaghter AC and MS modes, respectively) are constants (do not depend on $\tilde{\tau}$) while $A_0(\tilde{\tau})$ (the amplitude of the parent MRI) is given by $A_0(\tilde{\tau}) = A_+ e^{\tilde{\tau}} + A_- e^{-\tilde{\tau}}$. The latter reflects the multiplicity 2 of the corresponding MRI eigenvalue for $\gamma = 0$. However, as the amplitude of the MRI mode grows exponentially in time, the resonant interaction between the triad members starts to be effective and the various modes change accordingly. In addition, the growing MRI inevitably nonresonantly excites a zero frequency MS perturbation with amplitude $H_0(\tau) = A_0^2(\tau)$. Thus, the first terms on the right hand sides of equations (4) describe the independent linear eigen perturbations while the rest of the terms account for the influence of the resonant as well as nonresonant interactions. The goal therefore is to find the appropriate dynamical equations for the long-time evolution of the amplitudes of the initial eigen perturbations, namely, $A_0(\tilde{\tau})$, $A_1(\tilde{\tau})$ and $H_1(\tilde{\tau})$.

Following the discussion above, the asymptotic method for deriving the amplitude equations for the three - mode resonant system is generalized to near-resonance triads by allowing for a small mismatch, $\Delta\omega$. The present analysis is carried out for beta values that are slightly above the threshold of the first MRI parent mode. Thus, for a three - mode system with MRI zero frequency, $\omega_{1,-1} = 0$, fast ($l=1$) or slow ($l=-1$) AC frequency, $\omega_a = \omega_{k,l}$ that is calculated through equation (2) for $b = b_1$, and real MS frequency, $\omega_h$, the near-resonance condition may therefore be written as

$$\omega_h = \omega_a + \Delta\omega. \tag{5}$$

Here $\Delta\omega$ signifies an arbitrary frequency mismatch, i.e. the frequency difference of the daughter waves that may be excited in general case of widely arbitrary initial conditions.

## 2.2 Evolution equations for the amplitudes of detuned triads

As indicated above, $\Delta\omega$ is assumed to be of order of $\gamma$, and hence may be written as:

$$\Delta\omega = \gamma\Delta\widetilde{\omega}. \qquad (\Delta\widetilde{\omega}\sim\gamma^0). \tag{6}$$

Assuming now $A_0 \sim \gamma$, ($H_0 = A_0 \sim \gamma^2$), $A_1 \sim H_1 \sim \gamma^{3/2}$ the physically plausible evolution equations for the amplitudes of the participating eigenmodes are given by:

$$\gamma^2 \frac{d^2 A_0}{d\widetilde{\tau}^2} = \gamma^2 A_0 + E_{0,0} A_0 H_0 + E_{-1,1}[A_1^* H_1 \, e^{-i\widetilde{\tau}\Delta\widetilde{\omega}} + c.c.] + O(\gamma^4), \tag{7}$$

$$\gamma \frac{dA_1}{d\widetilde{\tau}} = i\, D_{0,1} A_0 H_1 \, e^{-i\widetilde{\tau}\Delta\widetilde{\omega}} + O(\gamma^{7/2}), \tag{8}$$

$$\gamma \frac{dH_1}{d\widetilde{\tau}} = i\, C_{0,1} A_0 A_1 \, e^{i\widetilde{\tau}\Delta\widetilde{\omega}} + O(\gamma^{7/2}). \tag{9}$$

The real non-linear coupling coefficients $E_{0,0}$, $E_{-1,1}$, $D_{0,1}$ and $C_{0,1}$ of zeroth order in $g$ in equations (7)-(9) are from Shtemler et al. (2014) [see also Table 1 below]. These coefficients are introduced in order to eliminate secular terms by satisfying the solvability conditions for the nonhomogeneous problems, which arise as perturbations to the homogeneous problems for the corresponding eigenmodes due to weakly nonlinear interaction between them.

Introducing now the following zeroth order (in $g$) scaled amplitudes:

$$a_0 = \frac{A_0}{\gamma \widehat{A}_0}, \qquad a_1 = \frac{A_1}{\gamma^{3/2} \widehat{A}_1}, \qquad h_1 = \frac{H_1}{\gamma^{3/2} \widehat{H}_1}, \tag{10}$$

where

$$\widehat{A}_0 = \sigma_D \frac{1}{\sqrt{|E_{0,0}|}}, \qquad \widehat{A}_1 = \frac{\Lambda^{3/2}}{\sqrt{|C_{0,1} E_{-1,1}|}}, \qquad \widehat{H}_1 = \frac{\Lambda^{3/2}}{\sqrt{|D_{0,1} E_{-1,1}|}}, \tag{11}$$

equations (7)-(9) are reduced to leading order in $\gamma$ to the following $\gamma$-free form [similar to Shtemler et al. (2014)]:

$$\frac{d^2 a_0}{d\theta^2} = \frac{1}{\Lambda^2}(a_0 - a_0^3) + \sigma_{DE}[a_1^* h_1 \, e^{-i\theta\,\delta\omega} + a_1 h_1^* e^{i\theta\,\delta\omega}], \tag{12}$$

$$\frac{da_1}{d\theta} = i a_0 h_1 \, e^{-i\theta\,\delta\omega}, \tag{13}$$

$$\frac{dh_1}{d\theta} = i\sigma_{CD} a_0 a_1 \, e^{i\theta\,\delta\omega}, \tag{14}$$

where

$$\theta = \Lambda\tilde{\tau}, \qquad \delta\omega = \frac{\Delta\tilde{\omega}}{\Lambda}, \qquad \Lambda = \sqrt{\left|\frac{D_{0,1}C_{0,1}}{E_{0,0}}\right|}, \qquad \sigma_{CD} = \sigma_C\sigma_D, \qquad \sigma_{DE} = \sigma_D\sigma_E, \qquad (15)$$

and

$$\sigma_C = sign(C_{0,1}), \qquad \sigma_D = sign(D_{0,1}), \qquad \sigma_E = sign(E_{-1,1}). \qquad (16)$$

The negative sign of the cubic term on the right hand side in equation (12) reflects the negative value of $E_{0,0} = -27/35$ [Liverts et al. (2012b)]. Note that system (12) - (14) depends only on two coupling parameters $\sigma_{CD}$ and $\sigma_{DE}$ instead of three, $\sigma_C$, $\sigma_D$ and $\sigma_E$ by introducing the factor $\sigma_D$ into the definition of $a_0$ in (10) - (11). Thus, the resulting resonant system is characterized by four independent parameters: $\Lambda^2$, $\delta\omega$ and $\sigma_{CD} = \pm 1$, $\sigma_{DE} = \pm 1$.

Equations (12)-(14) may be further simplified by introducing the absolute values and phases of the complex amplitudes of the daughter modes, $a_1(\theta)$, $h_1(\theta)$ (it is recalled that $a_0(\theta)$ is real-valued):

$$a_1(\theta) = \rho_a(\theta) e^{-i\phi_a(\theta)}, \qquad h_1(\theta) = \rho_h(\theta) e^{-i\phi_h(\theta)}. \qquad (17)$$

Substituting (17) into the (12) - (14) yields:

$$\frac{d^2 a_0}{d\theta^2} = \frac{1}{\Lambda^2}(a_0 - a_0^3) + F, \qquad (18)$$

$$\frac{d\rho_a}{d\theta} = -a_0 \rho_h \sin\phi, \qquad (19)$$

$$\frac{d\rho_h}{d\theta} = \sigma_{CD} a_0 \rho_a \sin\phi, \qquad (20)$$

$$\frac{d\phi}{d\theta} = \left(\sigma_{CD}\frac{\rho_a}{\rho_h} - \frac{\rho_h}{\rho_a}\right) a_0 \cos\phi - \delta\omega, \qquad (21)$$

where

$$\phi = \phi_a(\theta) - \phi_h(\theta) - \theta\delta\omega \qquad (22)$$

is the dynamical phase that characterizes the triad (the triad's dynamics does not depend seperately on each of the phases $\phi_a$ and $\phi_h$), and the term $F = 2\sigma_{DE}\rho_a\rho_h \cos\phi$ represents the driving force of the Duffing's oscillator. Equations (18) - (21) are complemented by the following initial conditions:

$$a_0(0) = a_{0,0}, \qquad \frac{da_0}{d\tilde{\tau}}(0) = \dot{a}_{0,0}, \qquad \rho_a(0) = \rho_{a,0}, \qquad \rho_h(0) = \rho_{h,0}, \qquad \phi(0) = \phi_0. \qquad (23)$$

Table 1 presents the coefficients of systems (7) - (9) and (18) – (21) for seven (out of infinitely many) possible resonant triads that involve the first MRI mode ($k = 1$, $l = -1$) as a parent mode [see also Table 3 in Shtemler et al. (2014)]. The triad stability criteria that emerge from Table 1 are that $\sigma_{CD} > 0$ is necessary and sufficient while $\sigma_{CD} < 0$ is necessary for instability. This wil be established explicitly in Section 2.3 and numerically in Section 3. As observed in Table 1 triads are stable if the role of the AC daughter wave is played by a fast AC mode, and unstable for slow AC daughter waves. Finaly note that unstable triads have typically negative values of $\sigma_{DE} = -1 < 0$.

Table 1. Eigen-frequencies $\omega_{k,l}(\beta_1)$ of the daughter AC modes and nonlinear resonant coupling coefficients for triads of MS and fast or slow AC daughter modes with the MRI parent mode ($k = 1, l = -1$) at its threshold beta $\beta_1$. Triads are named stable (unstable) depending on their behavior due to nonlinear interactions.

| Triad | Stable | Unstable | Stable | Unstable | Stable | Unstable | Stable |
|---|---|---|---|---|---|---|---|
| Axial wavenumber, $k$ | 1 | 2 | 2 | 3 | 3 | 4 | 4 |
| $\beta_k/\beta_1$ | 1 | 3 | 3 | 6 | 6 | 10 | 10 |
| Fast ($l=1$)/Slow ($l=-1$) AC daughter mode, $l$ | 1 | $-1$ | 1 | $-1$ | 1 | $-1$ | 1 |
| $\omega_{k,l}(\beta_1)$ | $\sqrt{7}$ | $\approx \sqrt{7/2}$ | $\approx \sqrt{31/2}$ | $\sqrt{10}$ | $\sqrt{27}$ | $\approx \sqrt{39/2}$ | $\approx \sqrt{83/2}$ |
| $C_{0,1}$ | $-0.68$ | $-3.40$ | $-5.06$ | $+2.01$ | $+1.81$ | $+16.46$ | $+13.81$ |
| $\sigma_C$ | $-1$ | $-1$ | $-1$ | $+1$ | $+1$ | $+1$ | $+1$ |
| $D_{0,1}$ | $-0.20$ | $+94.97$ | $-0.03$ | $-4.40$ | $+0.11$ | $-0.44$ | $+0.011$ |
| $\sigma_D$ | $-1$ | $+1$ | $-1$ | $-1$ | $+1$ | $-1$ | $+1$ |
| $E_{-1,1}$ | $-0.14$ | $-2.16$ | $+0.026$ | $+3.78$ | $+0.66$ | $+1.66$ | $-0.01$ |
| $\sigma_E$ | $-1$ | $-1$ | $+1$ | $+1$ | $+1$ | $+1$ | $-1$ |
| $\Lambda = \sqrt{\left|\frac{D_{0,1}C_{0,1}}{E_{0,0}}\right|}$ | 0.42 | 20.5 | 0.44 | 3.4 | 0.51 | 3.07 | 0.44 |
| $\sigma_{CD}$ | $+1$ | $-1$ | $+1$ | $-1$ | $+1$ | $-1$ | $+1$ |
| $\sigma_{DE}$ | $+1$ | $-1$ | $-1$ | $-1$ | $+1$ | $-1$ | $-1$ |

*Manley-Rowe relation. Differential relation for Duffing force and zero limit of the frequency mismatch*

Multiplying equations (19) and (20) for the amplitudes of the daughter modes by $\rho_a(\theta)$ and $\rho_h(\theta)$, respectively, and adding the results yield:

$$\sigma_{CD}\rho_a^2 + \rho_h^2 = \sigma_{CD}\rho_{a,0}^2 + \rho_{h,0}^2 = const. \tag{24}$$

The conservation law (24) is known as the Manley-Rowe relation.

Resonant triads with zero frequency-mismatch, $\delta\omega = 0$, have been qualitatively studied in [Shtemler et al. (2013), Shtemler et al. (2014)]. It has been shown for that case the force term in the equation for the parent MRI mode is identicaly constant ($F = F_0 \equiv 2\sigma_{DE}\rho_{a,0}\rho_{h,0}\cos\phi_0$), and the equation for the parent MRI mode decouples from those of the daughter modes. For non-zero mismatch however, this is non longer true and the rate of change of the driving force may be dervide from equations (19) - (21) and is given by:

$$\frac{dF}{d\theta} = 2\sigma_{DE}\,\delta\omega\,\rho_a\,\rho_h\,\sin\phi. \tag{25}$$

*Stable and unstable triads*

As is demonstrated in [Shtemler et al. (2013), Shtemler et al. (2014)] resonant triads with $\delta\omega = 0$ are divided into two sub-groups according to the signs of the nonlinear coupling coefficients: (i) stable ($\sigma_{CD} > 0$) and (ii) unstable ($\sigma_{CD} < 0$) daughter waves. This result is extended to the current case of non-zero frequency mismatch. This may be done by introducing the auxiliary dependent variable $\Theta(\theta) = \int_0^\theta a_0(\theta) \sin\phi(\theta)\, d\theta$. With that change of the independent variable the daughter waves equations (21) - (22) may be written as follows:

$$\frac{d^2 f}{d\Theta^2} - \kappa^2 f = 0, \qquad f = \{\rho_a, \rho_h\}, \qquad \kappa^2 = -\sigma_{CD}. \tag{26}$$

Consequently, the solutions for $f$ have the following form

$$f = C_+ \exp[\kappa\Theta(\theta)] + C_- \exp[-\kappa\Theta(\theta)], \tag{27}$$

so that the linearly stable daughter AC and MS waves may be nonlinearly destabilized if the following necessary condition

$$\kappa^2 = -\sigma_{CD} > 0, \qquad (\sigma_{CD} < 0) \tag{28}$$

is satisfied. The corresponding effective growth rates of the daughter waves may be defined in the following way:

$$\Gamma_\infty = \lim_{\theta \to \infty} |\Gamma(\theta)|, \qquad \Gamma(\theta) = \frac{1}{\theta}\int_0^\theta a_0(\theta) \sin\phi(\theta)\, d\theta, \tag{29}$$

where $\Gamma_\infty$ is the long-time limiting value of the instantaneous growth rate of the daughter waves. As will be seen in the next sections, the value of $\Gamma_\infty$ depends on the initial conditions, and triad that are characterized by $\sigma_{CD} < 0$ and $\Gamma_\infty = 0$ and therefore are stable in the sense described by equation (29) may still be prone to explosive instability if the initial amplitudes of triads are increased.

*The partial solutions for unstable triads*

As was indicated above, unstable triads are characterized by negative factor $\sigma_{CD} = -1 < 0$. As a result, for the subset of initial conditons that satisify $\rho_{a,0} = \rho_{h,0}$ the Manley-Rowe relation (24) for unstable triads yields:

$$\rho(\theta) = \rho_h(\theta) \equiv \rho_a(\theta). \tag{30}$$

Consequently, the two equations for the daughter wave amplitudes coalesce into the following single equation:

$$\frac{d\rho_a}{d\theta} = -a_0\, \rho_a\, \sin\phi. \tag{31}$$

## 2.3. Explosive solutions for unstable triads

Near-resonant unstable triads with $\sigma_{CD} = -1$ are now considered for the subset of perturbations that satisfy condition (30). In that case, system (18) - (21) acquires the following form:

$$\frac{d^2 a_0}{d\theta^2} = \frac{1}{\Lambda^2}(a_0 - a_0^3) + 2\sigma_{DE}\rho^2 \cos\phi, \tag{32}$$

$$\frac{d\rho}{d\theta} = -a_0\, \rho \sin\phi, \tag{33}$$

$$\frac{d\phi}{d\theta} = -2a_0\, \cos\phi - \delta\omega. \tag{34}$$

Numerical solutions of the above system indicate that explosive-type solutions in which the amplitudes of both the parent as well as the daughter modes tend to infinite values in a finite time ($\theta = \theta_e$) may exist (see Section 3). Guided by the numerical finite values of the triad's phase, the search for asymtotic explosive solutions starts by realizing that $a_0(\theta)\cos\phi(\theta)$ is finite close to the explosion time $\theta_e$. This occurs if $\cos\phi \to 0$ simultaneously with the explosive growth of $a_0(\theta)$ and $\rho_a(\theta)$, i.e. the limiting solution in the vicinity of $\theta_e$ is:

$$\phi(\theta) \cong \mp\frac{\pi}{2}(2M + 1) + \phi_e(\theta), \quad a_0(\theta) \cong a_e(\theta), \quad \rho(\theta) \cong \rho_e(\theta), \tag{35}$$

where $M = 0,1,2...$ is the number of the phase branch, and $\phi_e \to 0$, $a_e \to \infty$, $\rho_e \to \infty$, while the product $a_e(\theta)\phi_e(\theta)$ remains finite as $\theta \to \theta_e$.

Thus, in the limit $\theta \to \theta_e$

$$\cos\phi = \pm\sin\phi_e \cong \pm\phi_e, \quad \sin\phi = \mp\cos\phi_e \cong \mp 1. \tag{36}$$

Defining now the scaled variables:

$$\tilde{a}_e(\tilde{\theta}) = \theta_e\, a_e(\theta), \quad \tilde{\rho}_e(\tilde{\theta}) = \theta_e^2\sqrt{|\delta\omega|}\rho_e(\theta), \quad \tilde{\phi}_e(\tilde{\theta}) = \frac{1}{\theta_e\, \delta\omega}\phi_e(\theta), \quad \tilde{\theta} = \frac{\theta}{\theta_e}, \tag{37}$$

equations (32)-(34) in approaching the moment of explosion are reduced with the aid of (35)-(36) to the following self-similar form that is independent of $|\delta\omega|$ and $\theta_e$:

$$\frac{d^2\tilde{a}_e}{d\tilde{\theta}^2} = -\frac{1}{\Lambda^2}\tilde{a}_e^3 \pm 2\sigma_{DE}\sigma_{\delta\omega}\tilde{\rho}_e^2\,\tilde{\phi}_e, \tag{38}$$

$$\frac{d\tilde{\rho}_e}{d\tilde{\theta}} = \pm\tilde{a}_e\tilde{\rho}_e, \tag{39}$$

$$\frac{d\tilde{\phi}_e}{d\tilde{\theta}} = \mp 2\tilde{a}_e\tilde{\phi}_e - 1, \tag{40}$$

where $\sigma_{\delta\omega}$ is the sign of $\delta\omega$.

To further simplify the problem, the following relation between $\tilde{\phi}_e(\tilde{\theta})$ and $\tilde{a}_e(\tilde{\theta})$ is assumed:

$$\tilde{\phi}_e(\tilde{\theta}) = \frac{const}{\tilde{a}_e(\tilde{\theta})}. \tag{41}$$

Substituting (41) into (38)-(40) the asymptotic solution of the latter is given by:

$$\tilde{a}_e(\tilde{\theta}) = \frac{\tilde{K}_0}{1-\tilde{\theta}},$$

$$\tilde{\rho}_e(\tilde{\theta}) = \frac{\tilde{K}_1}{(1-\tilde{\theta})^2},$$

$$\tilde{\phi}_e(\tilde{\theta}) = \tilde{K}_2\,(1-\tilde{\theta})\,, \tag{42}$$

where

$$\tilde{K}_0 = \pm 2, \qquad \tilde{K}_1^2 = -\frac{6}{\sigma_{DE}\, \sigma_{\delta\omega}}\left(1+\frac{2}{\Lambda^2}\right) > 0, \qquad \tilde{K}_2 = -\frac{1}{3}. \qquad (43)$$

Returning to the original variables, the solution near the explosion time is described by:

$$a_e(\theta) = \frac{K_0}{\theta_e - \theta}, \qquad \rho_e(\theta) = \frac{K_1}{(\theta_e - \theta)^2}, \qquad \phi_e(\theta) = K_2\, (\theta_e - \theta)^{+1}, \qquad (44)$$

where $q_e$ is a free constant in the present asymptotic explosive solution

$$K_0 = \pm 2, \qquad K_1^2 = -\frac{6}{\sigma_{DE}\, \delta\omega}\left(1+\frac{2}{\Lambda^2}\right) > 0, \qquad K_2 = -\frac{1}{3}\delta\omega. \qquad (45)$$

In particular, solution (44)-(45) demonstrates that the amplitude of the parent MRI mode $a_e(\theta) \sim (\theta_e - \theta)^{-1}$ is less singular than that of the daughter modes $\rho_e(\theta) \sim (\theta_e - \theta)^{-2}$ in approaching the moment of explosion. It should be mentioned however that the explicit solution (44)-(45) is valid for $K_1^2 > 0$. This condition is satisfied for positive mismatches only since, as was indicated above, the factor $\sigma_{DE} = -1$ is negative for unstable triads (see Table 1). In that solution the value of the moment of explosion is indetermined as it naturaly depends on the initial conditions (see numerical simulations in the next section). It is further noticed that the amplitude of the parent mode is independent of the mismatch value, while the amplitude of the daughter modes tend to infinity as the mismatch vanishes. This means that the solution for the finite mismatch case does not converge to the solution obtained for the $\delta\omega = 0$ case.

Finaly note that the asymptotic explosive problem (38)–(40) depends on two parameters $\Lambda$ and $\sigma_{DE}\sigma_{\delta\omega}$ (the latter may acquire the values $\pm 1$). It determines a universal explosive solution that is independent of mismatch value and the value of the explosion time, i.e. self-similar in $|\delta\omega|$ and $q_e$. In this connection the positivity of the mismatch value ($\delta\omega > 0$) in the exact explosive solution (44)-(45) is a corollary of the restrictive assumption (41), while the general explosive problem (38)–(40) that is based on the less restrictive relations (35) - (37) is valid for both kinds of triads with positive and negative mismatches.

It is seen in Figure 1 that after an initial transient period the numerical solutions of problem (18) - (23), plotted in the self-similar variables (37) for positive mismatch values, are very close to the asymptotic explosive ones (44) - (45) in the vicinity of the explosion time. As is further seen in Figure 1 the two curves for two different mismatch values: $\delta\omega = +1$, and $\delta\omega = +2.5$, overlap as the explosion time is approached for $0.9 \lesssim \tilde{\theta} = \theta/\theta_e < 1$. In addition, the powers of $(\theta_e - \theta)$ as well as the amplitude factors $K_0$, $K_1$ and $K_2$ in relations (44)-(45) are found to be very close to the numerically obtained limiting values at vanishing $(\theta_e - \theta)$. Although explicit explosive solution is unavailable for negative mismatch values, numerical solutions of equations (38) -(40) indicate that such solutions do exist and exhibit the same self-similar characteristics as those obtained analytically for positive mismatch values. Indeed, the amplitudes of the three participating modes that are depicted in Figure 2 for two different mismatch values: $\delta\omega = -1$, and $\delta\omega = -2.5$ overlapp as $\tilde{q}$ approaches the explosion time ($\tilde{q} = 1$).

As was indicated above, negativity of $\sigma_{CD}$ is the necessary condition for the triad instability. This is also a necessary condition for the existence of explosive solutions, since otherwise for $\sigma_{CD} > 0$ the Manley-Row conservation law guarantees the boundedness of the triad's amplitudes. With $\sigma_{CD} < 0$ explosive solutions do satisfy the Manley-Row conservation law.

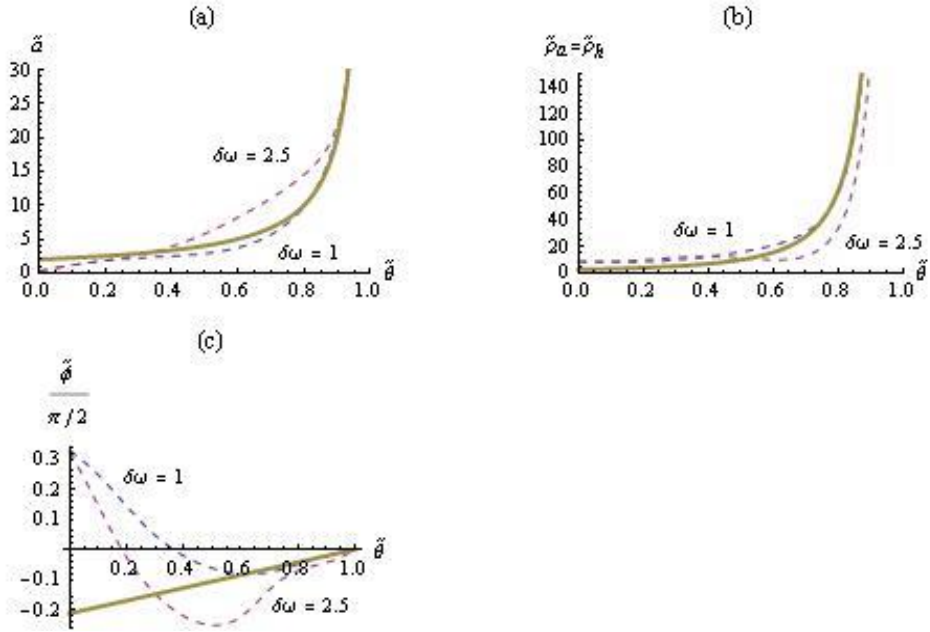

Figure 1. Comparision numerical explosive solutions with explicit one for amplitudes and dynamic phases of unstable triads with positive mismatches; $\sigma_{CD} = -1$, $\sigma_{DE} = -1$, $k = 2, l = -1$, $\Lambda = 20.5$ for initial conditions (46). Solid lines for explicit explosive solution in the form (42) –(43); dashed lines for two numerical solutions for $\delta\omega = +1$, $(\theta_e \approx 4.56)$ and $\delta\omega = +2.5$, $(\theta_e \approx 4.63)$, respectively. The subscript e in $\tilde{a}_e$ and $\tilde{\phi}_e$ in Figure 1 (a) is omitted.

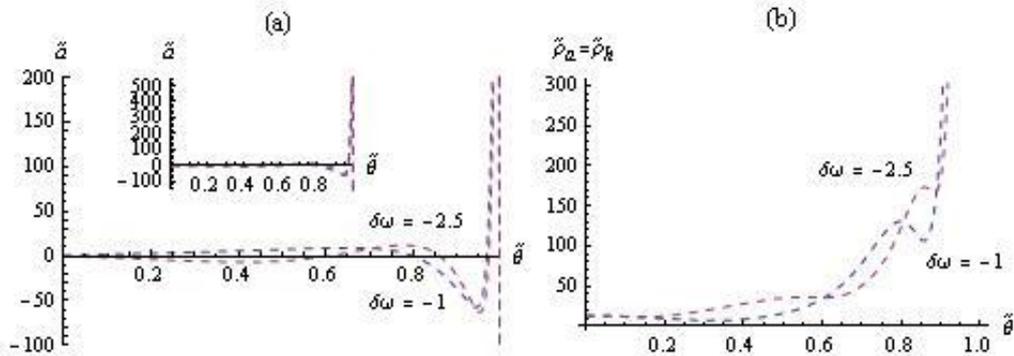

Figure 2. Numerical solutions for amplitudes and dynamic phase in self-similar variables (40) for two explosive unstable triads with $\delta\omega < 0$ with initial conditions (46); $\delta\omega = -1$, $(\theta_e \approx 5.59)$ and $\delta\omega = -2.5$, $(\theta_e \approx 5.76)$, $\sigma_{CD} = -1$, $\sigma_{DE} = -1$, $k = 2, l = -1$, $\Lambda = 20.5$.
To distinguish two curves outside the vicinity of the maximum value $\tilde{a}_e \approx 550$ the top of Figure 2(a) is cut off (completely shown in insertion in Figure 2(a)). The subscript e in $\tilde{a}_e$ in Figure 2(a) is omitted.

# 3. RESULTS OF NUMERICAL SIMULATIONS

As was mentioned above the near-resonance triads are divided into two sub-groups according to the signs of the nonlinear coupling coefficients in Table 1: (i) unstable ($\sigma_{CD} < 0$) and (ii) stable ($\sigma_{CD} > 0$) triads. In addition, the unstable triads are further divided into two classes for $\delta\omega > 0$ and $\delta\omega < 0$. The dynamical behavior of near-resonat unstable triads may dramatically differ from that of the their strict resonance counterparts. The nonlinear saturation of the parent MRI-mode and exponential-like nonlinear instability of the daughter waves (two linearly stable waves, namely, slow AC and MS eigenmodes) that occur in the long–time limit for $\delta\omega = 0$ is replaced for $\delta\omega \neq 0$ by an explosive growth of all three modes.

The various types of the response of the unstable and stable triads with both positive and negative mismatches are shown in Figures 3-14. All simulations are carried out for the first two triads (first two colomns in the Table 1) one of which corresponds to stable and the second to unstable triad which involve fast and slow AC daughter waves, respectively. The results for the rest of the triads are not presented here since they are in qualitative agreement with first two (stable and unstable, respectively).

Most simulations of the present study are carried out for the following initial conditions (if not mentioned otherwise):

$$a_{0,0} = 0.05, \dot{a}_{0,0} = 0.5, \rho_0 = 0.4, \phi_0 = \frac{\pi}{4}. \tag{46}$$

## 3.1 Unstable triads.

*Numerical simulations for unstable triads.*

Results of typical numerical simulations for unstable triads with positive and negative mismatch are presented in Figures 3 – 4 and Figures 5 – 7, respectively. They are compared with the results for the same system with zero mismatch. For positive mismatch, in qualitative agreement with the explicit explosive solution (44)-(45), the amplitudes of all three modes (as well as the Duffing's oscillator force *F*) grow monotonically as the moment of explosion is approached. Increasing the positive mismatch from $\delta\omega = 1$ (Figure 3) to $\delta\omega = 10$ (Figure 4), the solution preserves its explosive nature, but the explosion time is delayed from $\theta_e \approx 4.56$ up to $\theta_e \approx 8.4$. For such large values of the mismatch the explosive instability amplitudes of daughter waves are of the same order of magnitude as their zero-mismatch exponentially growing counterparts (Figure 4). This is in contrast to the small mismatch case for which the explosive amplitudes of daughter waves are much bigger than exponentially growing ones in the zero-mismatch case (Figure 3). Simultaneously for both $\delta\omega = 1$ and $\delta\omega = 10$ amplitudes of the parent modes explosively grow and are much bigger than finite amplitudes in the zero-mismatch case.

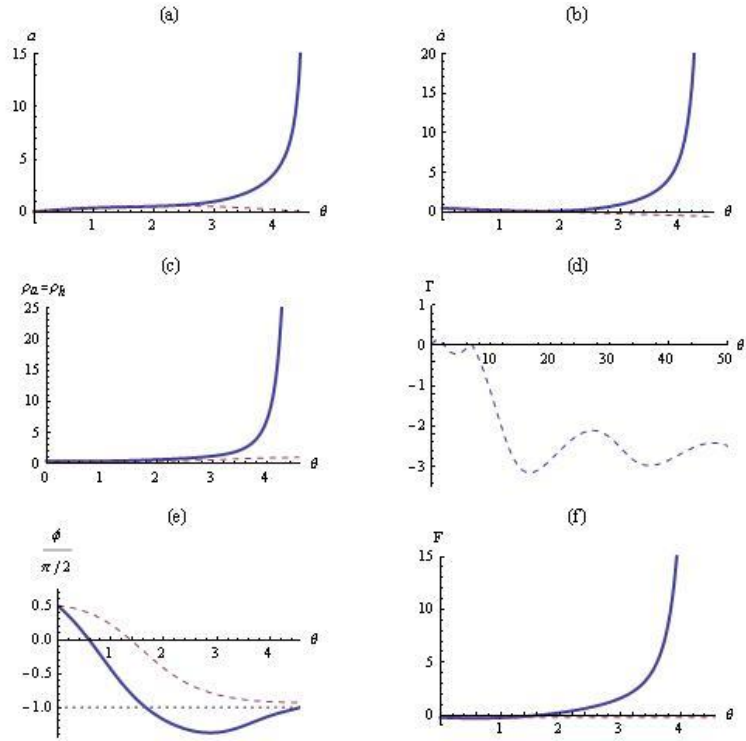

Figure 3. Amplitudes and dynamic phase of unstable triads with
$\sigma_{CD} = -1, \sigma_{DE} = -1, k = 2, l = -1, \Lambda = 20.5$ and initial conditions (46).
Solid lines for $\delta\omega = +1, (\theta_e \approx 4.56)$, dashed lines for $\delta\omega = 0$. Figure 1(d) illustrates the exponential-like long-time growth rate of daughter waves. The subscript 0 in $a_0$ in Figures 3 (a),( b) is omitted.

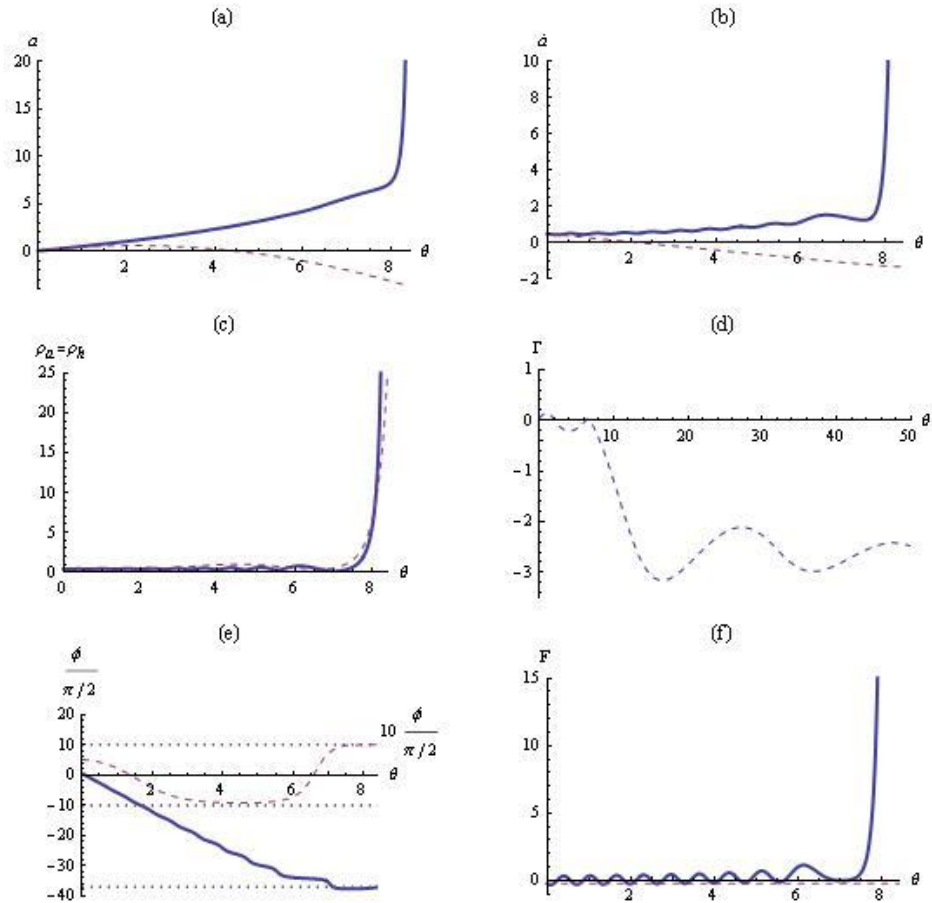

Figure 4. Amplitudes and dynamic phase of unstable triads as in Figure 3; $\delta\omega = +10, (\theta_e \approx 8.4)$.

Turning now to unstable near-resonance triads with negative mismatches $\delta\omega < 0$, they also demonstrate explosive growth of all three amplitudes as well as of the Duffing's oscillator force. Although in both cases ($\delta\omega < 0$ and $\delta\omega > 0$) explosion occurs when the phase reaches one of the values: $\phi \to \mp\frac{\pi}{2}(2M + 1)$, ($M = 0,1,2 ...$) predicted by the relations (35), the case $\delta\omega < 0$ looks somewhat more complicated than that for $\delta\omega > 0$. Figure 5 shows that the amplitudes of triads with $\delta\omega = 0$ grow in approaching to the moment of explosion non-monotonically as distinct from the $\delta\omega > 0$ case. For $\delta\omega < 0$ the explosive growth is preceeded by one or more local extremums (see Figures 5 and 7). Another specifics of the $\delta\omega < 0$ case is the phase change by several jump-like variations prior to the moment of explosion. With increasing of the absolute value of the negative mismatch, the moment of explosion sharply grows, so that for the standart initial conditions (46) the explosive instability disappears altogether starting from $\delta\omega \approx -3.93$ (this value depends on the initial conditions). This lack of explosive solutions for large negative $\delta\omega$ (see Figure 6 for $\delta\omega \approx -4$) occurs due to the degeneration of the local growth rate $\Gamma(\theta)$ in (29). Namely, the departure $\Gamma(\theta)$ in (29) from zero is of an oscillating nature (Figure 6) and consequently the relation $\Gamma(\theta) \approx 0$ is satisfied on the average. The limiting growth rate $\Gamma_\infty$ vanishes in that case, which signifies the stability of the triads. The dependence of the stability properties of triads with $\delta\omega < 0$ on the initial conditions is illustrated in Figures 6 and 7. Changing the low initial amplitudes of the triad's members in Figure 6 to higher values in Figure 7 while keeping $\delta\omega \approx -4$ leads to change from the bounded oscillating triads with $\Gamma_\infty = 0$ in the former case to explosively growing triads with $\Gamma_\infty > 0$ in the latter.

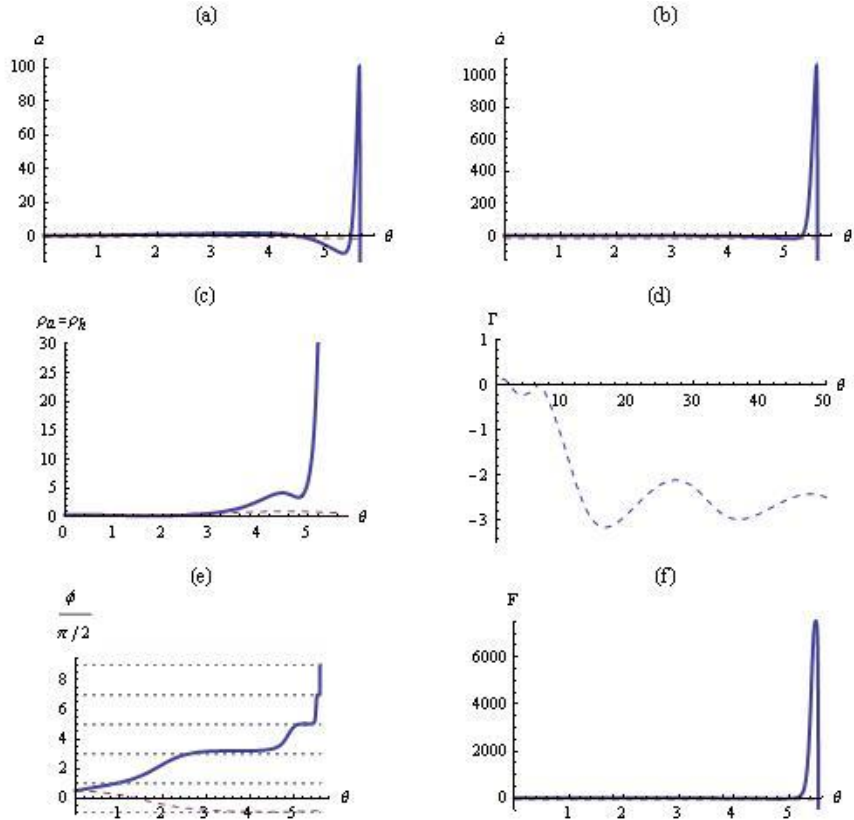

Figure 5. Amplitudes and dynamic phase of unstable triads as in Figure 3; $\delta\omega = -1$, ($\theta_e \approx 5.59$).

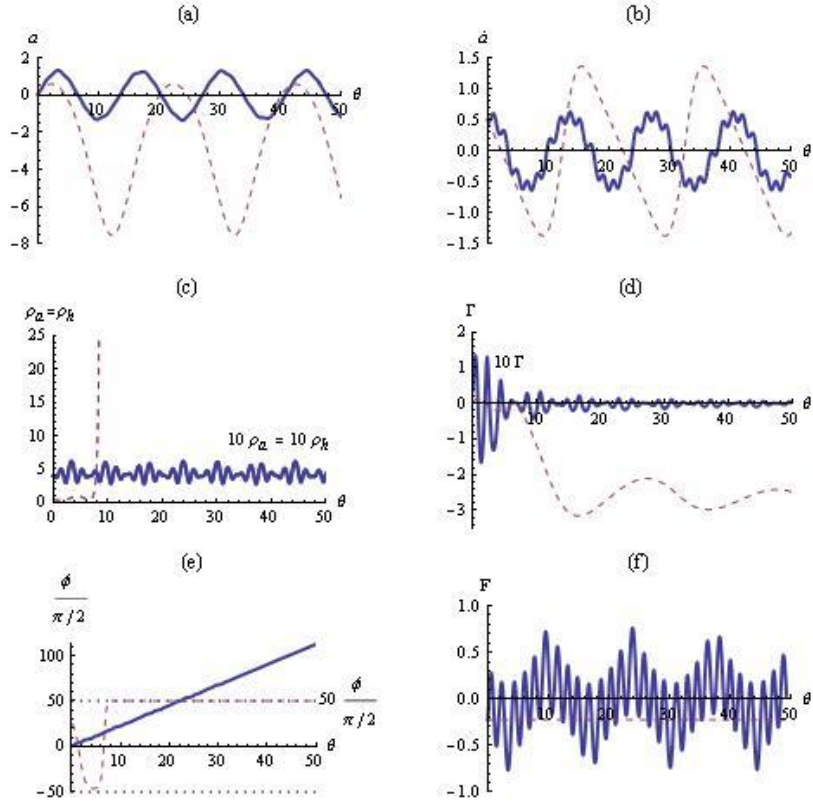

Figure 6. Amplitudes and dynamic phase for triads with $\sigma_{CD} = -1, \sigma_{DE} = -1, k = 2, l = -1, \Lambda = 20.5$ and initial conditions (46). Solid lines for $\delta\omega = -4$, dashed lines for $\delta\omega = 0$.

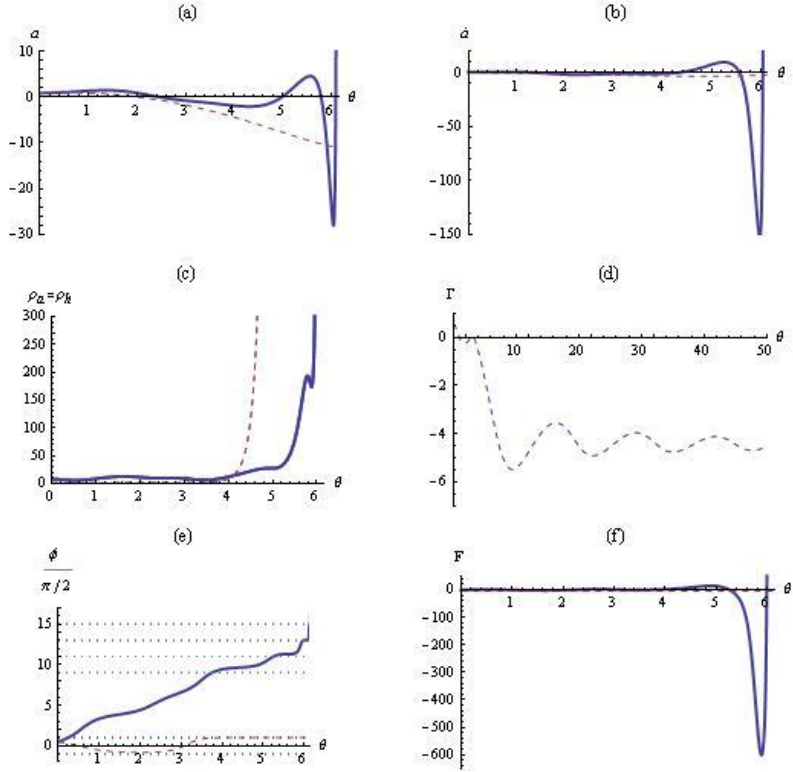

Figure 7. Amplitudes and dynamic phase for the triads with $\sigma_{CD} = -1, \sigma_{DE} = -1, k = 2, l = -1$, $\Lambda = 20.5$ and higher than in (46) initial amplitudes: $a_{0,0} = 0.8, \rho_{a,0} = \rho_{h,0} = 0.8$.
Solid lines for $\delta\omega = -4$, ($\theta_e \approx 6.16$), dashed lines for $\delta\omega = 0$.

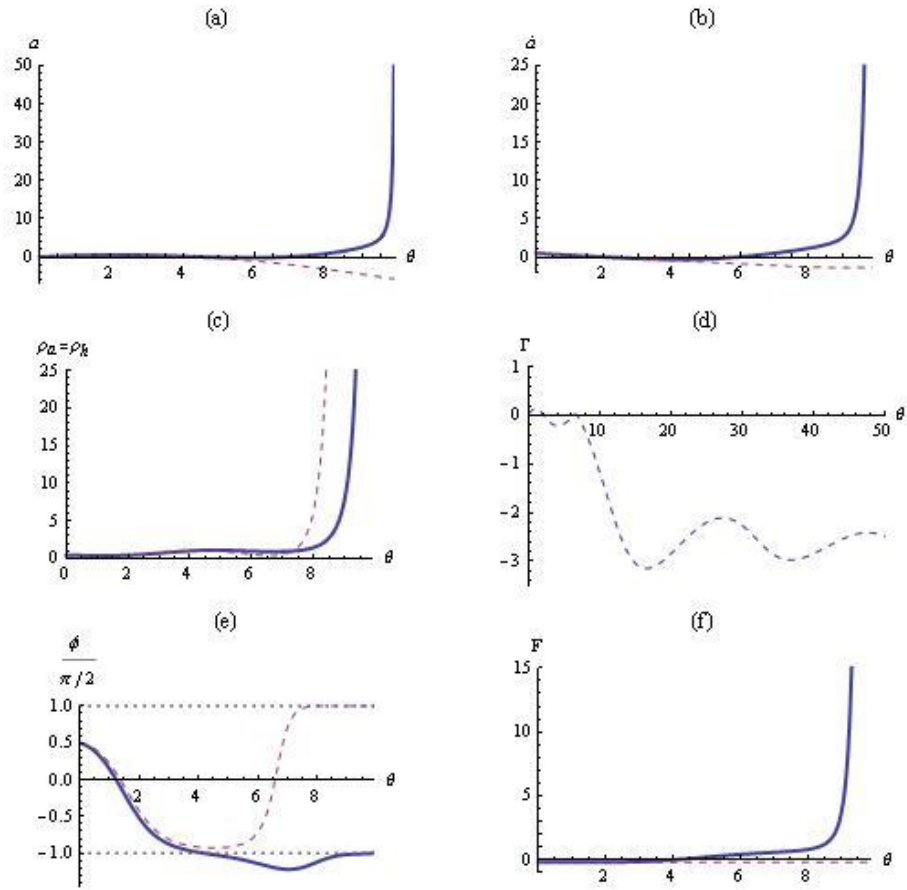

Figure 8. Amplitudes and dynamic phase of unstable triads as in Figure 3; $\delta\omega = +0.1$, ($\theta_e \approx 9.91$).

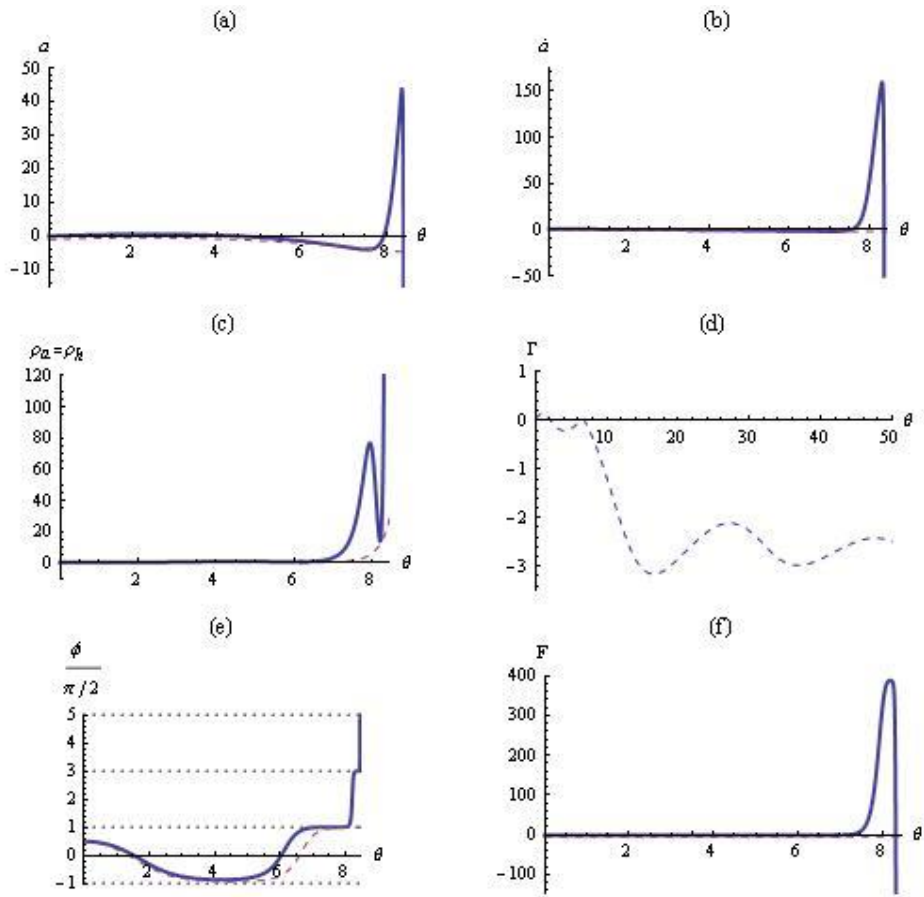

Figure 9. Amplitudes and dynamic phase of unstable triads as in Figure 5, $\delta\omega = -0.1$, ($\theta_e \approx 8.45$).

*Small mismatch values*

As numerical simulations demonstrate for small values of mismatches (both positive and negative), the amplitudes and dynamic phases are qualitatively similar to those for finite values of mismatches. However, they are not close neither one to another nor to those for strict resonance case (see Figures 8-9 for $\delta\omega = +0.1$ and $\delta\omega = -0.1$).

Thus, as distincts from positive-mismatch triads in Keplerian discs that are explosively unstable independently of initial conditions, for negative- mismatch triads the explosive instability occurs when the initial amplitudes of the triads exceed thresholds growing with the mismatch module value. Such behavior for negative- mismatch is in a qualitatively with that in infinite plasmas [Rosenbluth et al. (1969)].

*Dependence on initial conditions. Competition of strict- and near-resonance triads.*

Initial conditions play a critical role in the competition between exponential-like and explosive instabilities of the daughter waves, as is indeed observed by simulations for the strict - and near - resonance triads, respectively. Figure 10(c) with Figure 3(c) present the results for a fixed mismatch value with different initial conditions for the dynamical phase $\phi_0 = 3\pi/4$ and $\phi_0 = \pi/4$. They illustrate the competition of the explosive and exponential-like instabilities of the daughter waves. Thus, for $\phi_0 = \pi/4$ the explosive instability for $\delta\omega = 1$ dominates the exponential-like instability for $\delta\omega = 0$ (Figure 3(c)), and for $\phi_0 = 3\pi/4$ the exponential-like instability dominates the explosive one (Figure 10(c)). In both cases the amplitudes of daughter waves may grow up to very large values at which the the weakly-nonlinear analysis loses its validity. The same phenomenon of instabilities' competition but now depending on initial conditions for $\dot{a}_0 = \dot{a}_{0,0}$ is demonstrated by Figure 5(c) and Figure 11(c) with $\dot{a}_{0,0} = 0.5$ and $\dot{a}_{0,0} = 0.25$, respectively. In that case ($\delta\omega = -1$) for $\dot{a}_{0,0} = 0.5$ the explosive instability of the daughter waves dominates the exponential-like one with $\delta\omega = 0$ (Figure 5(c)), while $\dot{a}_{0,0} = 0.25$ is the transition initial value at which exponential-like instability is generated practicaly simultaneously with the explosive instability.

Based on the results of the simulations it can be conjectured that for the case of a slow AC daughter wave, near-resonance unstable triads ($\sigma_{CD} < 0$) may explosively grow either independently of initial conditions or depending on initial conditions for positive and negative mismatch, respectively. Also, since the parent MRI mode can unboundedly grow only for the case of detuned triads the competition is relevant only for the daughter waves.

In the present examples of explosive triads the values of the explosion time depends on the initial conditions. The explosive interaction may take place even for sufficiently small initial amplitudes of all waves for positive mismatch values (e.g. see Figure 12), however the value of the moment of explosion grows significantly in that case.

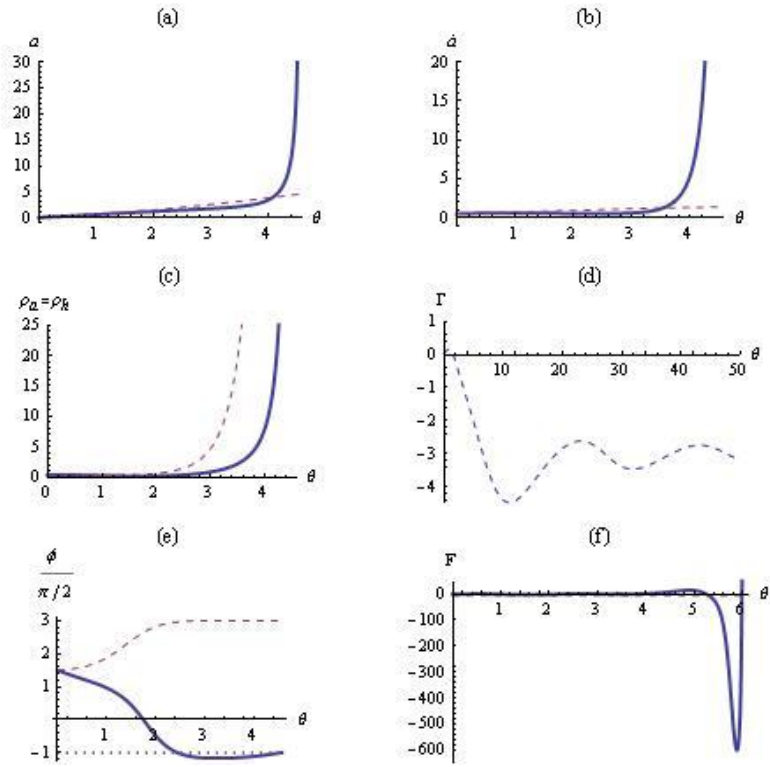

Figure 10. Amplitudes and dynamic phase of unstable triads as in Figure 3, $\delta\omega = 1$, ($\theta_e \approx 4.57$) with initial dynamic phase $\phi_0 = \frac{3\pi}{4}$.

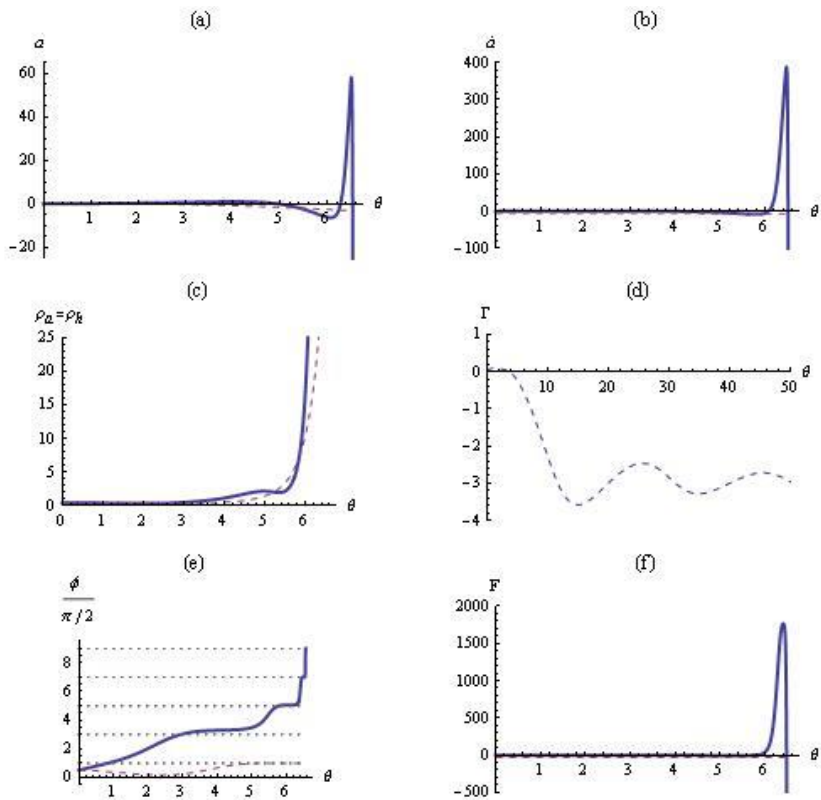

Figure 11. Amplitudes and dynamic phase of unstable triads as in Figure 5, $\delta\omega = -1$, ($\theta_e \approx 6.57$) with the initial derivative $\dot{a}_{0,0} = 0.25$.

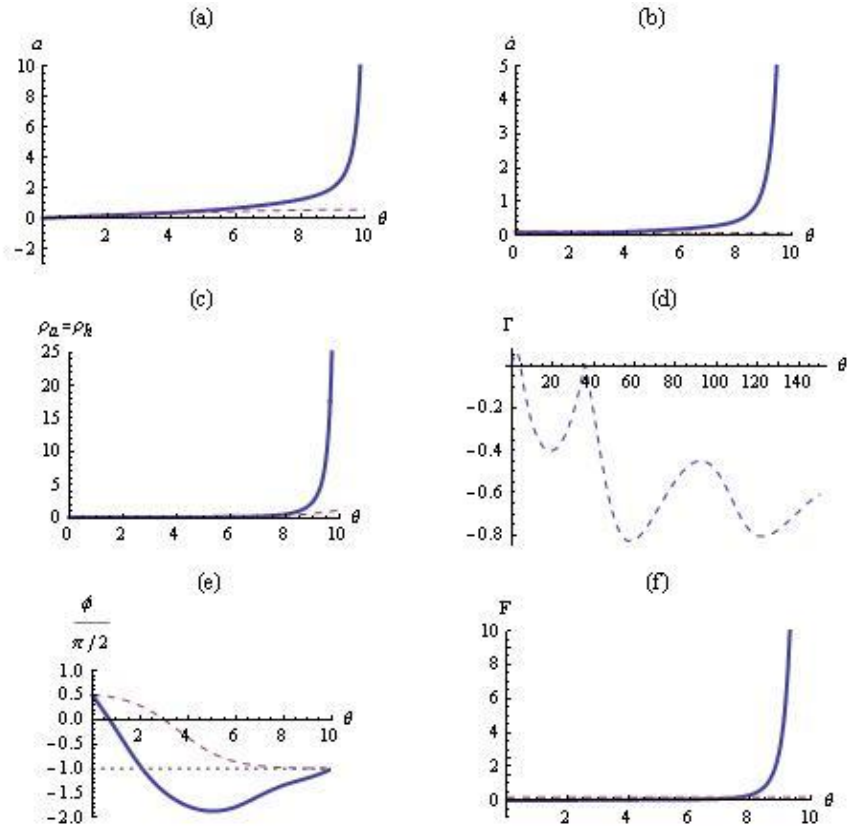

Figure 12. Amplitudes and dynamic phase of unstable triads as in Figure 3, $\delta\omega = +1$, $(\theta_e \approx 10.0)$ with the lower than in (46) initial amplitudes: $a_{0,0} = 0.01$, $\dot{a}_{0,0} = 0.1$, $\rho_{a,0} = \rho_{h,0} = 0.08$.

### 3.2 Stable near-resonance triads.

The above picture is completely changed in the case of stable near-resonance triads. In particular, the Duffing's oscillator force that unboundedly grows with time for the unstable triads oscillates with a finite amplitude in the stable case. Stable triads with both positive and negative mismatches preserve qualitatively the same behavior as in the case of zero mismatch (Figures 13-14). Some difference is observed in a somewhat less regular behavior of the amplitudes of the parent MRI and the daughter waves as compared with strict-resonance system.

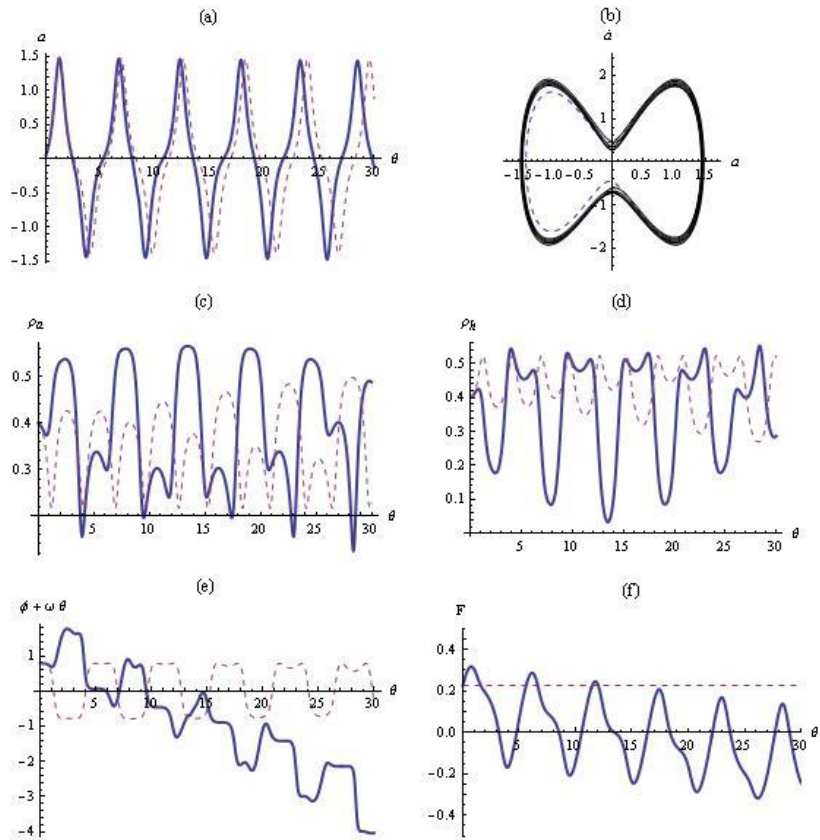

Figure 13. Amplitudes and dynamic phase for the stable triad with
$\sigma_{CD} = -1, \ \sigma_{DE} = -1, k = 1, l = -1, \Lambda = 0.42$ and initial conditions (46).
Solid lines for $\delta\omega = +1$, dashed lines for $\delta\omega = 0$.

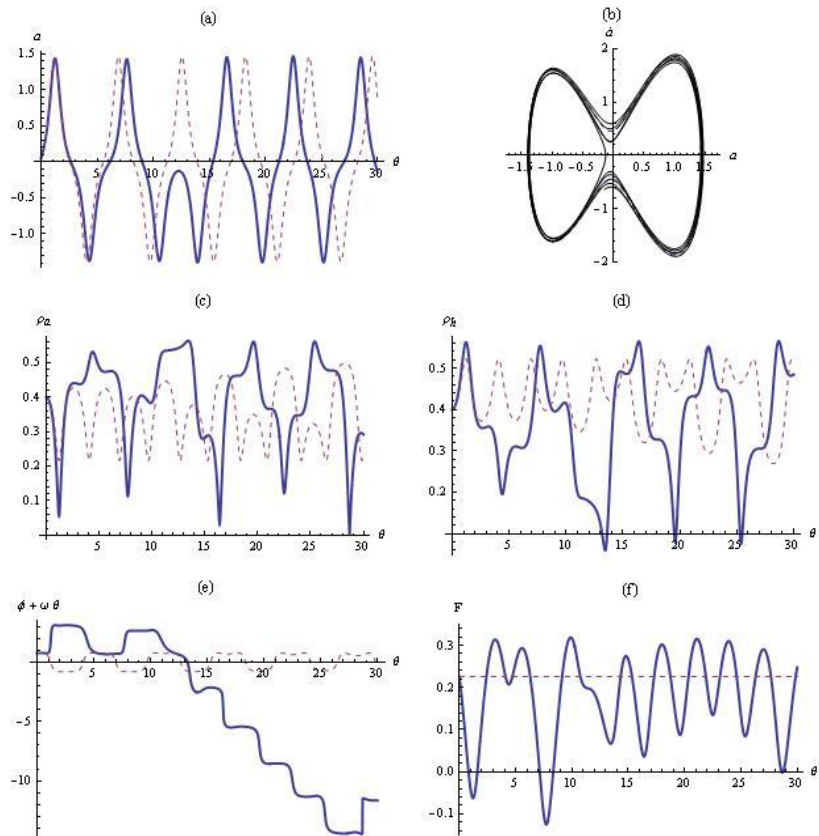

Figure 14. Amplitudes and dynamic phase of the stable triad as Figure 13.
Solid lines for $\delta\omega = -1$, dashed lines for $\delta\omega = 0$.

## 4. Summary and Discussion

The dynamics of near-resonance triads with a small nonzero mismatch $\delta\omega$ is investigated, and is shown to dramatically deviate from the nonlinear behavior of the strict-resonance triads. In particular, it is demonstrated that an explosive magneto-rotational instability of near-resonant unstable triads may occur.

If the AC daughter wave is a fast AC mode, the amplitudes of all three resonantly interacting modes with $\delta\omega \neq 0$ remain bounded, as in the case of $\delta\omega = 0$. However, the presence of a slow AC daughter wave with $\delta\omega \neq 0$ dramatically changes the nonlinear behavior of the unstable strict-resonance wave triads. The nonlinear-saturation of the parent MRI-mode and the nonlinear exponential-like growth of the daughter waves (a linearly stable pair of *slow* AC and MS eigenmodes) that occurs for $\delta\omega = 0$ is replaced by an explosive growth during a finite time of all those three modes. The necessary condition for the existence of explosive solutions for has been derived $\delta\omega \neq 0$ that is coincide with that for the nonlinear exponential-like instability for $\delta\omega = 0$. The nonlinear scenario of the explosive instability depends however on the sign of the frequency mismatch. Thus, while positive values of the latter result in an explosive instability regardless of the initial conditions, negative values lead to exlplosive instability only beyond some threshold of the initial amplitudes. The explosive solutions demonstrate that the amplitude of the parent mode is less singular than those of the daughter modes in approaching the moment of explosion. As is demonstrated above by comparison of strict- and near-resonance systems, the value of the small mismatch is of principal significance for explosion instability in thin Keplerian discs. It is clear from the results presented above that this regime of instability is realized exclusively due to the non-constancy of the force term in the Duffing equation for the parent mode (the force term is constant for the zero mismatch case). It is the deviation of the Duffing's oscillator force from the zero-mismatch constant value that provides the positive feedback that leads to the explosive growth of the parent modes that in turn amplifies the growth of the dautghter waves. Paradoxically, explosively unstable near- resonance triads may grow much faster than their strict-resonance counterparts. Although in the long-time limit the explosive instability of daughter waves will always dominate the exponential-like one, for certain initial conditions and during finite time that preceded the moment of explosion, the exponential-like growth of the strictly coupled triads may occur earlier than the explosive growth in near- resonance coupling. The competition between the two scenarios of instability is determined by the initial conditions.

Although the results are presented for first two modes (first two colomns in the Table 1) one of which correspond to stable and the second to unstable triad, the results for other possible triads are in qualitative agreement with the first two (stable and unstable, respectively), the only such resulting values as the moment of explosion, characteristic amplitudes etc. are varied with the choice of the triad.